\documentclass[sigconf]{acmart}

\usepackage{booktabs}
\usepackage{listings}
\usepackage{multirow}
\usepackage{threeparttablex}
 \usepackage{subfig}
\usepackage{enumitem}
\usepackage{todonotes}

\usepackage[many]{tcolorbox}  %

 \usepackage{setspace}    %
\definecolor{main}{HTML}{5989cf}    %
\definecolor{sub}{HTML}{cde4ff}     %

\tcbset{
    sharp corners,
    colback = white,
    before skip = 0.1cm, %
    after skip = 0.1cm  %
}      

\newtcolorbox{boxB}{
    enhanced,
    boxrule = 0pt,
    colback = sub,
    borderline west = {1pt}{0pt}{main}, 
    borderline west = {0.75pt}{2pt}{main}, 
    borderline east = {1pt}{0pt}{main}, 
    borderline east = {0.75pt}{2pt}{main}
}

\usepackage{tabularx}
\AtBeginDocument{%
  }

\setcopyright{acmlicensed}
\copyrightyear{2026}
\acmYear{2026}
\acmDOI{XXXXXXX.XXXXXXX}
\acmConference[ICSE 2026]{48th International Conference on Software Engineering}{April 2026}{Rio de Janeiro, Brazil}
\acmISBN{978-1-4503-XXXX-X/2018/06}

\begin{document}

\title{Understanding Code Agent Behaviour: An Empirical Study of Success and Failure Trajectories}

\author{Oorja Majgaonkar}
\affiliation{%
  \institution{University College London}
  \city{London}
  \country{United Kingdom}}
\email{oorja.majgaonkar.24@ucl.ac.uk}

\author{Zhiwei Fei}
\affiliation{%
  \institution{Nanjing University}
  \city{Nanjing}
  \country{China}}
\email{zhiweifei@smail.nju.edu.cn}

\author{Xiang Li}
\affiliation{%
  \institution{University College London}
  \city{London}
  \country{United Kingdom}}
\email{x.li.25@ucl.ac.uk}

\author{Federica Sarro}
\affiliation{%
  \institution{University College London}
  \city{London}
  \country{United Kingdom}}
\email{f.sarro@ucl.ac.uk}

\author{He Ye}
\authornote{Corresponding author.}
\affiliation{%
  \institution{University College London}
  \city{London}
  \country{United Kingdom}}
\email{he.ye@ucl.ac.uk}

\renewcommand{\shortauthors}{Oorja Majgaonkar et al.}

\begin{abstract}
The increasing deployment of Large Language Model (LLM) agents for complex software engineering tasks has created a need to understand their problem-solving behaviours beyond simple success metrics. While these agents demonstrate impressive capabilities in automated issue resolution, their decision-making processes remain largely opaque. This paper presents an empirical study of agent trajectories, namely the execution traces capturing the steps agents take when attempting to resolve software issues. We analyse trajectories from three state-of-the-art code agents (OpenHands, SWE-agent, and Prometheus) on the SWE-Bench benchmark, examining both successful and failed attempts.
Our investigation reveals several key insights into agent behaviour. First, we identify how distinct problem-solving strategies, such as defensive programming and context gathering, enable success in different scenarios. Second, we find that failed trajectories are consistently longer and exhibit higher variance than successful ones, with failure patterns differing significantly between agents. Third, our fault localisation analysis shows that while most trajectories correctly identify problematic files (72-81\% even in failures), success depends more on achieving approximate rather than exact code modifications.
These and other findings unveiled by our study, provide a foundation for understanding agent behaviour through trajectory analysis, contributing to the development of more robust and interpretable autonomous software engineering systems.

\end{abstract}



\setcopyright{none}
\settopmatter{printacmref=false}

\maketitle
\section{Introduction}

Large Language Models (LLMs) have rapidly advanced from simple code completion to complex repository-level problem solving \cite{rewardrepair, selfapr, chen_prometheus_2025, wang_2025_openhands, yang_swe-agent_2024, zhang_autocoderover_2024, xia_agentless_2024, bouzenia_repairagent_2025}. This shift has driven the emergence of code agents—systems that combine LLMs with scaffolding, external tools, and reasoning loops to autonomously resolve software engineering tasks. Benchmarks such as SWE-Bench \cite{swebench_verified} demonstrate that these agents can fix real-world GitHub issues, suggesting significant potential for automating software maintenance at scale. Despite their impressive capabilities, current evaluations treat agents as black boxes, focusing primarily on success rates while overlooking the rich information contained in their execution trajectories \cite{bouzenia_understanding_2025, wang_are_2025}.

The opacity of agent systems presents a challenge for practitioners and researchers for improving these systems. As they rely largely on experimental trial-and-error to improve these systems, it remains unclear whether failures stem from insufficient context, flawed reasoning, poor fault localization, or other factors. Trajectories, or the detailed logs of every step an agent takes during problem solving, offer an unexplored window into agent behaviour. A few studies~\cite{kim_prospector_2024, song_agentbank_2024, lin_evolution_2025} have identified the potential of trajectories to agent performance, such as through contrastive learning, while other work~\cite{bouzenia_understanding_2025, ceka_understanding_2025,chen_unveiling_2025} focuses on high level patterns and decomposing trajectory structure. However, they do not take advantage of trajectories to gain a detailed understanding of how agents approach tasks, where they struggle, and why they fail, beyond what they achieve.

Our study addresses this gap. We conduct the first comparative analysis of agent trajectories on the SWE-Bench benchmark, using three diverse systems: OpenHands \cite{wang_2025_openhands}, SWE-agent \cite{yang_swe-agent_2024}, and Prometheus \cite{chen_prometheus_2025}. Our investigation moves beyond binary success metrics to examine the pathways agents follow, comparing successful and failed attempts to identify critical factors that determine outcomes.

We investigate three research questions that progressively deepen our understanding of agent behaviour. First, we examine the unique capabilities of different agents by analysing issues that only specific agents can resolve, conducting deep dives into the trajectories to reveal problem-solving strategies. Second, we quantify the structural differences between successful and failed trajectories, uncovering patterns in trajectory length and variance that correlate with failure modes. Third, we investigate fault localisation capabilities at multiple granularities, assessing how precisely agents identify problematic code and how this precision relates to ultimate success.

By moving beyond binary success rates, this study offers a foundation for interpreting and improving code agents through trajectory-level understanding. We argue that robustness and interpretability—rather than raw leaderboard scores—are the key challenges in building the next-generation of autonomous software engineering systems.

In summary, our work makes the following contributions: 
\begin{itemize}
    \item \textbf{Trajectory dataset construction}: We assemble and normalise execution traces from three state-of-the-art code agents on SWE-Bench Lite and Verified.
    \item \textbf{Comparative trajectory analysis:} We provide the first detailed comparative qualitative study of trajectories across multiple state-of-the-art agents, revealing how successful problem-solving strategies as well as failure modes vary between systems.
    \item \textbf{Structural characterisation of failures:} We quantify differences in trajectory lengths, showing systematic patterns between success and failure.
    \item \textbf{Empirical study of fault localisation} We provide the first cross-agent analysis of localisation accuracy at file, function, and hunk levels.
    \item \textbf{Takeaways on agent behaviour:} We identify eight takeaways on agent behaviour, unveiling learnings on trajectory length, strategies, success and failure patterns, and fault localisation ability. 
\end{itemize}

\section{Related Work}\label{sec:background}
\subsection{Code Agents}

The challenge of complex, multi-step problems has led to the development of \emph{code agents} that move beyond stand-alone LLMs \cite{wang_agents_2024, ye2025adverintent, ye2024iter, he2023precisebugcollector, luo2024fine}. An agent is a system composed of one or more core LLMs enhanced by a \textit{scaffolding} of abstracts and tooling \cite{agents_anthropic}. This architecture enables an agent to perform complex reasoning \cite{wei_chain--thought_2022}, execute actions \cite{wang_executable_2024}, understand their outcomes, and plan its next activity \cite{huang_planning_2024}. They are therefore significantly more capable at problems requiring deeper understanding and multiple steps. Notable academic examples of software engineering agents include SWE-agent \cite{yang_swe-agent_2024}, AutoCodeRover \cite{zhang_autocoderover_2024}, OpenHands \cite{wang_2025_openhands}, RepairAgent \cite{bouzenia_repairagent_2025} and Prometheus \cite{chen_prometheus_2025}, as well as the comparable Agentless approach \cite{xia_agentless_2024}.

Agentic systems, with their diverse architectures combining different LLMs, frameworks, and reasoning strategies, are largely treated as opaque systems \cite{martinez_dissecting_2025}. The lack of systematic ablative studies and the inherent randomness of their core LLMs hinder our understanding of why they succeed or fail. Motivated by this gap, our work aims to systematically investigate agent behaviour to inform more robust and effective designs.

\subsection{Agent Trajectory Studies} 

Agent trajectories have proven useful for improving agent performance, such as using them in contrastive learning, fine-tuning or iterative optimisation \cite{kim_prospector_2024, song_agentbank_2024, lin_evolution_2025}. However, the value of using them for \textit{understanding} agents has been overlooked, despite  usage in industry \cite{agentevals, meng2025docent}.
To our knowledge, there are five works that look at this potential.

A quantitative study of top-performing agents on SWE-bench by \citeauthor{chen_unveiling_2025} revealed that Python execution errors during a task are a primary challenge, corresponding to increased reasoning overhead and lower success rates \cite{chen_unveiling_2025}. Another study of execution traces by \citeauthor{ceka_understanding_2025} highlights that fault reproduction is a weak link and finds that agent-generated patches are typically more minimal, localised, and less context-aware than human-written ones, especially for more difficult problems \cite{ceka_understanding_2025}, however, this study focuses more on the patch produced rather than the process. It has also been found that as agentic systems grow in complexity, new failure modes related to coordination and alignment overhead emerge \cite{cemri_why_2025}. Addressing the complexity of these execution trajectories, SeaView \cite{bula_seaview_2025} provides a tool for visually inspecting them to aid in the analysis. Finally, \citeauthor{bouzenia_understanding_2025} created a taxonomy and framework for trajectory patterns to achieve a high-level understanding, using RepairAgent, AutoCodeRover, and OpenHands as subjects \cite{bouzenia_understanding_2025}. 

While crucial for defining overarching patterns, current studies lack a fine-grained, bottom-up analysis of agent behaviour, particularly missing insights on partial progress that can be found by examining failing trajectories. Our work addresses this gap with a detailed comparative empirical study of code agent trajectories.

\section{Empirical Study Design}\label{sec:study_design}

We design our empirical study in three parts, starting with case studies of uniquely solved issues, followed by computing and comparing trajectory lengths and fault localisation ability. 

\subsection{Data Collection}
To understand the trajectories from multiple code agents, our work considers three code agents selected based on the criteria that they must: 1) be open source, 2) have trajectories available, and 3) be released within the past year and currently maintained.

We choose OpenHands \cite{wang_2025_openhands} and SWE-agent \cite{yang_swe-agent_2024} as they represent established state-of-the-art baselines in autonomous software engineering research, and Prometheus \cite{chen_prometheus_2025} as a novel methodology specialised in codebase-level understanding. We used their results on both SWE-bench Lite and Verified \cite{swebench_verified}. The trajectories for OpenHands and SWE-agent are collected from a public bucket made available by the SWE-Bench benchmark maintainers \cite{swebench_experiments}. Prometheus logs are collected directly from the experimental evaluation. The details of each agent are summarised in Table \ref{tab:agents}.

\begin{table}[tb]
    \centering
    \small
    \renewcommand{\arraystretch}{0.85}

    \caption{Selected Code Agents on Two SWE-bench Datasets}
    \label{tab:agents}
    \begin{tabular}{lllr}
        \toprule
        SWE-Bench & Agent name & Base LLM & Evaluation date \\
        \midrule
        \multirow{3}{*}{Lite} & OpenHands & CodeAct 2.1 Sonnet & 25/10/2024 \\
        & SWEAgent & Claude 4 Sonnet & 26/05/2025 \\
        & Prometheus & GPT-4o & 26/07/2025 \\
        \midrule
        \multirow{3}{*}{Verified} & OpenHands & CodeAct 2.1 Sonnet & 29/10/2024 \\
        & SWEAgent & Claude 4 Sonnet & 22/05/2025 \\
        & Prometheus & Devstral Medium 2507 & 11/08/2025  \\
        \bottomrule
    \end{tabular}
\end{table}
\subsection{Research Questions}
\textbf{RQ1: What bugs can the three agents uniquely resolve?}: We identify all those bugs that each agent is able to solve uniquely. We then conduct deep dives into samples from each of these unique sets, dividing the trajectory events into five categories for consistent comparison between trajectories.

\textbf{RQ2: How different are the failing and successful trajectories?}: We measure this by the number of steps in the trajectory, comparing the findings between successful and failed trajectories. 

\textbf{RQ3: To what extent do code agents succeed in locating faults before producing the patch?}: We calculate how precisely the agent's patches match the gold patch for the issue, so as to explore the agent's ability to understand the repository structure, progress in solving the issue, and localise the fault.

\subsection{Methodology for RQ1}

The SWE-Bench experiments public repository contains information on which issues were successfully resolved in each submission. This data is extracted for each of the agents studied to compute the following evaluation metrics.

The \textbf{uniqueness} of each agent's contribution is defined as the \textbf{set of issues resolved exclusively by that agent}. This is calculated using the set difference between an agent's set of resolved issues and the union of the other agents' sets. We do this for each agent studied.

After obtaining the unique issue sets for the three agents, we conduct a manual inspection of the trajectories in these sets to understand what enables one agent to succeed compared to the other two. Three trajectories are uniformly selected from each of the three unique sets as a representative sample and examined. 
As agent trajectories are long and complex, a structured analysis framework is developed to extract the main decision points. Each trajectory is segmented into five distinct phases of the problem-solving process:
\begin{enumerate}[leftmargin=*]
    \item Diagnosis: Initial identification of relevant files, classes, and functions.
    \item Problem Reproduction: Attempt to trigger the bug and observe the reported behaviour.
    \item Hypothesis and Plan: Articulated reasoning for a potential fix.
    \item Implementation: The first significant code modification made.
    \item Verification: Process of testing the solution to confirm the fix.
    
\end{enumerate}
In the case where the agent used multiple attempts to solve the issue, such as using newly learned information to change the plan, the trajectory was considered as a loop with the same five steps, with the verification step leading to a success which ends the attempt or a failure which triggers a new loop. We focus especially on hypothesis and implementation and look for points where failed approaches diverge from successful ones. 

This framework enables a consistent comparison across trajectories and can be used to determine divergence points in agent behaviour. 

\subsection{Methodology for RQ2}
We hypothesise that failed attempts contain more iterations than successful ones, as the agent likely gets stuck and goes down unsuccessful paths during the reasoning process. We compute the \textbf{number of steps} in each trajectory and then derive the statistical distribution of this value in failed and successful trajectories. 

Taking inspiration from the thought-action-result structure proposed by \citeauthor{bouzenia_understanding_2025} \cite{bouzenia_understanding_2025}, we define a \textit{step} as an event in the trajectory that is one of these three categories. Their work calculates length in terms of \textit{iterations}, which is such a sequence of thought, action, and result events. We instead opt to count these steps individually, so as to preserve granularity and potential for more detailed study and avoid making direct associations between adjacent events in the case they are not actually related. 

A Thought Step indicates the agent printing its reasoning or reflection. An Action Step can be any of tool calls, commands, or action-oriented interaction between the internal components, such as making a call to knowledge graph or communication between agents in a multi-agent system. A Result Step is an event containing the output or consequence of an Action. 

A trajectory is thus an array of Steps of one of these types. We implement parsers for each of the agents to convert the trajectory data into this unified format. Table \ref{tab:parser_summary_auto} describes how steps are extracted from each agent to give a unified interface over disparate agent output formats.

\begin{table*}[tb]
\centering
\small
\caption{Summary of Trajectory Parser Implementations for Different Agent Artifacts}
\label{tab:parser_summary_auto}
\renewcommand{\arraystretch}{1.0}
\begin{tabularx}{\textwidth}{|l|X|X|X|X|}
\hline
\textbf{Agent} & \textbf{Input Artifact} & \textbf{Thought Extraction} & \textbf{Action Extraction} & \textbf{Result Pairing} \\
\hline
\textbf{OpenHands} &
  JSON arrays of trajectory steps &
  From \texttt{assistant} text fields &
  Structured \texttt{tool\_calls} &
  By matching the \texttt{tool\_call\_id} of an action with its result \\
\hline
\textbf{SWE-agent} &
  JSON object with a \texttt{trajectory} list &
  From explicit \texttt{thought} and \texttt{response} fields &
  Structured \texttt{tool\_calls}; falls back to inferred commands when structure is absent &
  By matching \texttt{tool\_call\_id}s or, if unavailable, by attaching subsequent observations \\
\hline
\textbf{Prometheus} &
  Raw text log file (\texttt{.log}) &
  Parsed reasoning segments directly from log lines &
  Parsed \texttt{tool\_calls} arrays found within log messages &
  By associating command or patch outputs with the most recent preceding action \\
\hline
\end{tabularx}
\end{table*}

\subsection{Methodology for RQ3}
RQ3 explores the ability of the agents to accurately localise the fault and to what level of detail they succeed, beyond the final success and failure outcomes. We derive \textbf{three levels of fault localisation} measure how much progress the agent makes in identifying the location of the problematic code: correct file, correct function, and correct hunk or lines. This value can be correlated with the outcome. For every issue, we compare the agent's candidate patch and the gold patch available in the benchmark dataset. We extract the modified files, modified functions, and modified hunks from each, and check whether all the entities in the gold patch are present in the candidate patch. Additionally, we implement a ``lenient'' hunk match check, which allows line differences within a hunk up to a tolerance of 5, represented as a partial match using the value 0.5. Note that this check still requires all hunks themselves to be present, and only allows for deviations in line numbers within that hunk, so missing hunks would lead to an outcome of 0. Thus, each patch pair has a result object defined by the format in Table \ref{tab:faultloc_format}.

\begin{table}[tb]
    \centering
    \small
\renewcommand{\arraystretch}{0.8}

\caption{Fault Localisation Result Format}
\label{tab:faultloc_format}

    \begin{tabularx}{\linewidth}{llX}\toprule
         Field name& Possible values&Description\\\midrule
         file\_match& 1, 0&Candidate targeted all gold files (1) or not (0)\\
 function\_match& 1, 0&Candidate targeted all gold functions (1) or not (0)\\
 hunk\_match& 1, 0.5, 0&Candidate targeted all gold hunks in full (1), with minor line differences (0.5), or not (0) \\ \bottomrule\end{tabularx}
    
\end{table}
These results are then tallied to understand trends across the complete dataset in two ways:
\begin{itemize}
    \item \textit{Marginal frequencies}: We count number of successes and failures for each metric individually.
    \item \textit{Joint frequencies}: We compute the frequency of each unique combination of metric outcomes.
\end{itemize}
We expect to see a funnel-like pattern, as increasingly fewer patches are able to accurately locate the fault as the granularity increases.

We note that the relation between a perfect fault localisation as per this criteria and successful outcome is not trivial, as success on the benchmark is based on whether the patch passed all test cases. The presence of additional modifications in the candidate patch is ignored, as this question only seeks to know whether the agent could correctly identify where the fault was, even though it is possible that extra changes cause the overall solution to fail. Conversely, it is likely that many successful trajectories do not modify the same hunks as the gold patch, instead finding a different solution that still works. Thus, the gold patch should be considered a guide for comparison rather than an arbiter of success, but we occasionally refer to it as "correct" for simplicity. 

\section{Empirical Study Results}\label{sec:result}

\subsection{RQ1: Uniquely Repaired Issues}

\begin{figure}[tb]
\centering
\subfloat[Lite]{
\includegraphics[width=0.25\textwidth]{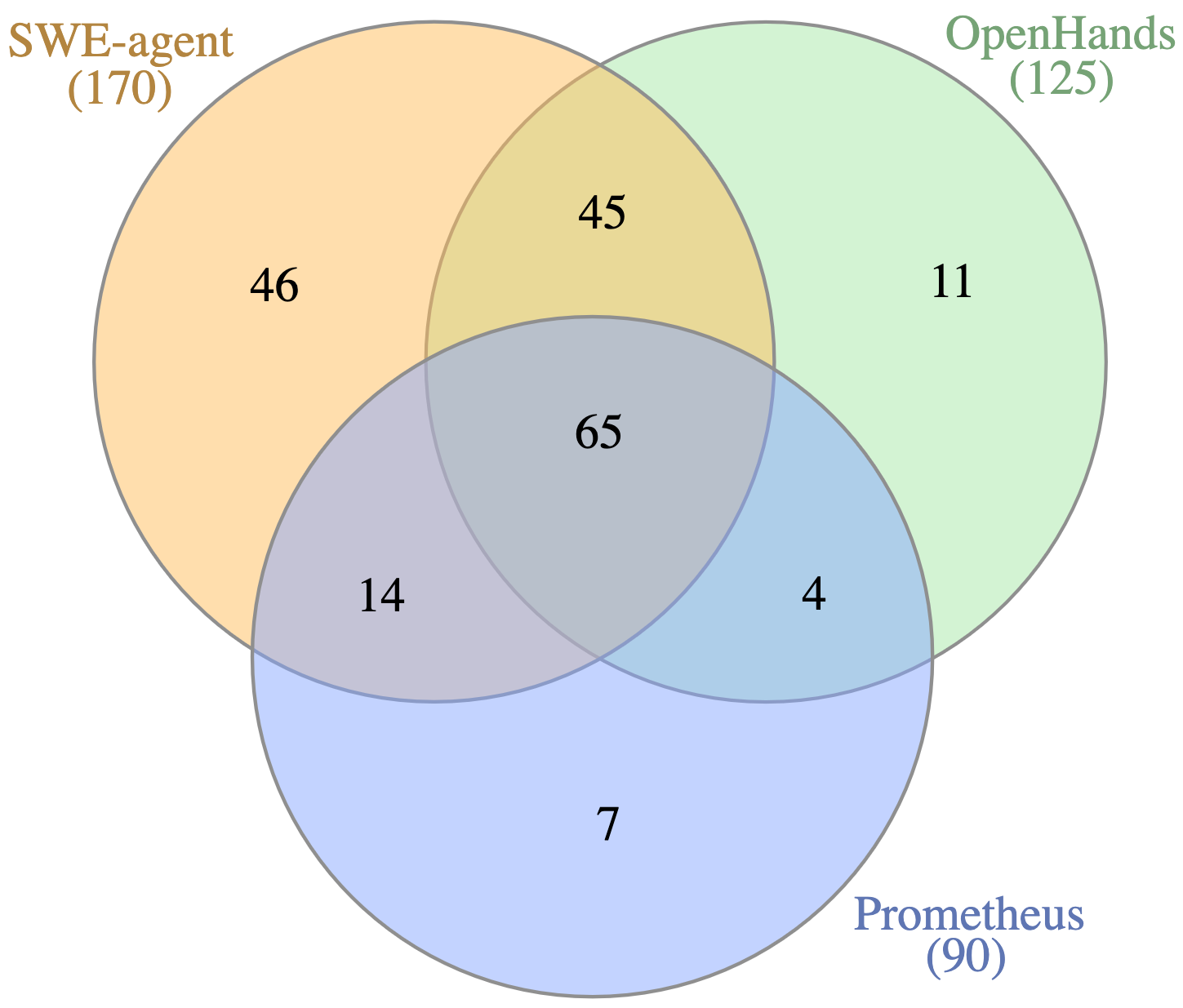}
\label{fig:1}
}
\subfloat[Verified]{
\includegraphics[width=0.25\textwidth]{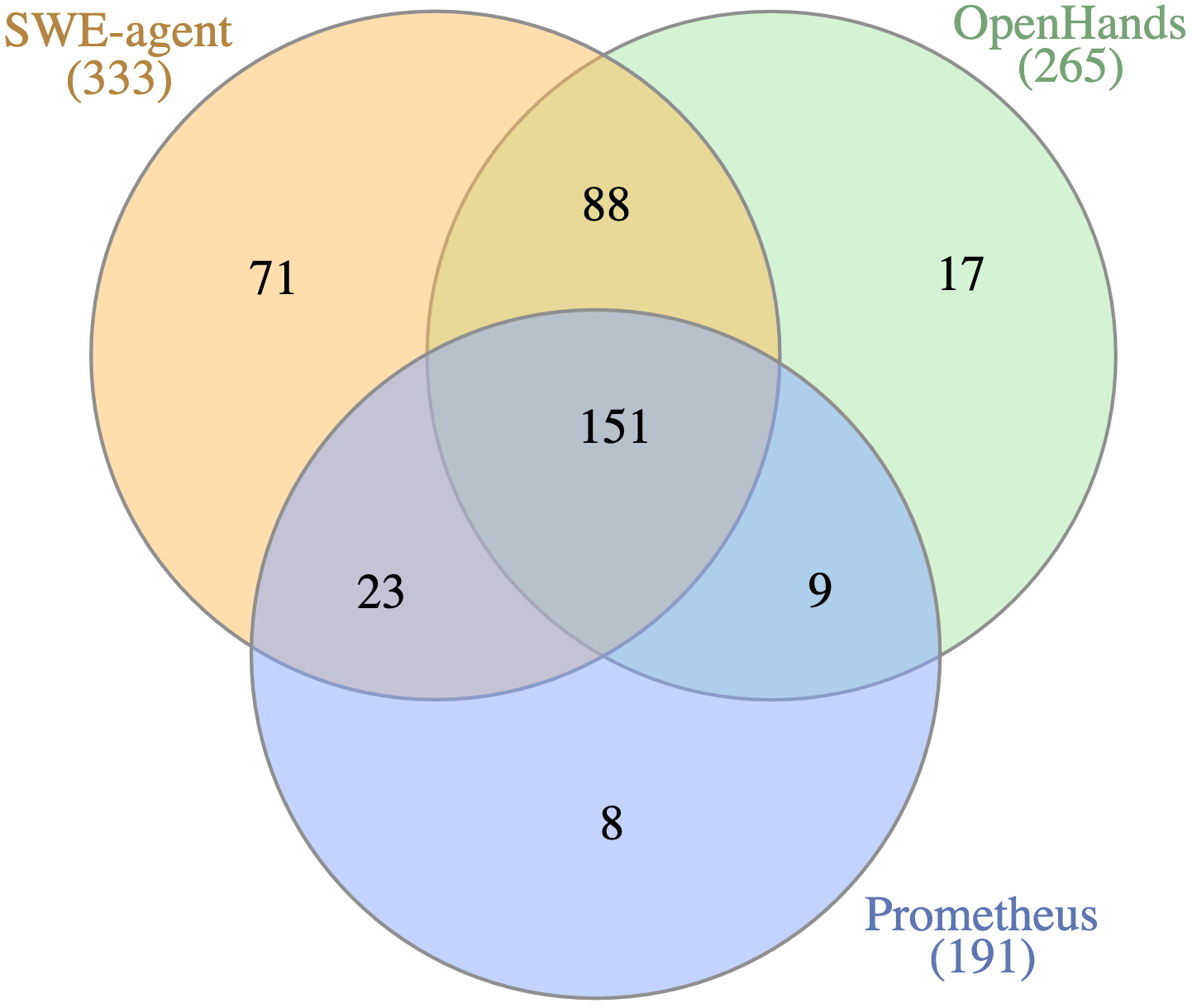}
\label{fig:2}
}
\caption{Resolved issues in SWE-Bench.}
\label{fig:swe-venn}
\end{figure}
    
Figures \ref{fig:swe-venn} illustrate the unique issues that each agent was able to solve in SWE-Bench Lite and Verified, respectively. We present three case studies, each diving into one issue from each of the unique sets of three agents, chosen uniformly at random.

\subsubsection{Prometheus uniqueness} Prometheus could solve 7 unique issues from SWE-Bench Lite and 8 in Verified. This limited improvement contrasts with the larger gains achieved by the other two agents on Verified.

\textbf{Case study: django-11964.} The issue description states that unexpected types are returned when creating Django model instances with TextChoices or IntegerChoices fields. 

Prometheus first conducted an analysis of the issue and identified that the problem was fundamentally about enum representation rather than database field behaviour. It arrived at a solution that overrode the \lstinline|__str__| method in both IntegerChoices and TextChoices classes, and modified \lstinline|IntegerField.to_python()| to handle enum instances. This approach was minimal and leveraged Django's existing patterns. The final patch addressed both the enum behavior and field conversion aspects of the problem.

SweAgent correctly reproduced the bug, but then went down a complex path of metaclass modification, attempting to change how enum values are stored during class creation rather than simply overriding instance method behaviour. 

OpenHands also reproduced the bug but misidentified the problem scope. It could not recognise this as an enum representation issue, instead creating a new descriptor method for the enum class database and field level, trying to modify ORM query behaviour and field validation logic. This led to increasingly complex, misguided solution attempts that never addressed the actual root cause.

SweAgent and OpenHands trajectories display a misunderstanding of the problem domain, whereas Prometheus displayed an understanding of the typical way of working in the Django repository. 

\begin{boxB} \textbf{Finding 1: Repository-aware context gathering enables knowledge of existing architectural patterns, which could be useful in localising the right abstraction level and understanding intended behaviour.} \end{boxB}

\subsubsection{OpenHands uniqueness} OpenHands could solve 11 unique issues in SWE-Bench Lite and 17 issues in Verified.

\textbf{Case study: pylint-6258.} The issue description states that the ignore flags were not being respected when Pylint was run in recursive mode. 

OpenHands formed the hypothesis that the \lstinline|_discover_files| method in pylinter.py was responsible for finding files when the --recursive=y flag was used, reasoning that this method traversed the file system and collected Python files before any ignore patterns were applied. The ignore logic was being executed too late, after the file list had already been discovered. It applied one large code change, which changed the relevant method from static to instance, so that linter instance's configuration could be accessed. Compared to the gold patch, the OpenHands patch achieved the right functionality, but was less efficient as it traverses directories to check for ignored files even when the root should be ignored. The gold patch additionally reveals that the same logic existed in a different function, and extracts it into a helper to avoid duplication.

SWE-Agent successfully reproduced the bug and performed the right reasoning about the problem at hand. However, its first code modification resulted in all files being ignored, which was a regression. The main point of deviation was its failure to convert \lstinline|_discover_files| to an instance method. As a result, SWE-Agent spent time debugging and creating new test cases, and could not access the config. 

Prometheus failed to reproduce the bug as its attempts hit the recursion limit. It was nevertheless able to formulate a hypothesis using static analysis, correctly identifying that filtering logic was being applied too late, and showed a plan to modify the function to accept ignore configurations. However, it then engaged in extensive context gathering as it tried to figure out how configuration options worked in the program, hitting another limit before finding a result, and thus failing to implement a patch. 
\begin{boxB} \textbf{Finding 2: Limiting recursion is not enough to prevent going down incorrect paths without avoiding recursive calls. Agents need mechanisms to abandon unproductive reasoning loops. }\end{boxB}

\begin{table*}[tb]
\centering
\small
\renewcommand{\arraystretch}{1.1}
\caption{Number of Steps in Successful and Failed Trajectories}
\label{tab:trajectory_stats}
\begin{tabular}{lrrrrrrrrrrrr}
\toprule
& \multicolumn{6}{c}{\textbf{SWE-Bench Lite}} & \multicolumn{6}{c}{\textbf{SWE-Bench Verified}} \\
\cmidrule(lr){2-7} \cmidrule(lr){8-13}
\textbf{Metric} & \multicolumn{2}{c}{\textbf{OpenHands}} & \multicolumn{2}{c}{\textbf{SWE-agent}} & \multicolumn{2}{c}{\textbf{Prometheus}} & \multicolumn{2}{c}{\textbf{OpenHands}} & \multicolumn{2}{c}{\textbf{SWE-agent}} & \multicolumn{2}{c}{\textbf{Prometheus}} \\
\cmidrule(lr){2-3} \cmidrule(lr){4-5} \cmidrule(lr){6-7} \cmidrule(lr){8-9} \cmidrule(lr){10-11} \cmidrule(lr){12-13}
& \textbf{Success} & \textbf{Failure} & \textbf{Success} & \textbf{Failure} & \textbf{Success} & \textbf{Failure} & \textbf{Success} & \textbf{Failure} & \textbf{Success} & \textbf{Failure} & \textbf{Success} & \textbf{Failure} \\
\midrule
\quad Avg & 61.01 & 79.90 & 158.36 & 178.31 & 87.16 & 136.47 & 54.05 & 98.62 & 151.95 & 180.10 & 146.59 & 220.92 \\
\quad Min & 15 & 12 & 60 & 73 & 25 & 33 & 18 & 24 & 58 & 66 & 28 & 11 \\
\quad Max & 297 & 296 & 537 & 351 & 192 & 691 & 295 & 297 & 447 & 377 & 546 & 715 \\
\quad Std Dev & 46.60 & 60.63 & 54.91 & 58.42 & 34.62 & 100.79 & 40.42 & 80.81 & 56.14 & 59.00 & 73.93 & 133.70 \\
\bottomrule
\end{tabular}
\end{table*}

\subsubsection{SWE-agent uniqueness} SWE-agent solved 46 and 71 unique issues in SWE-Bench Lite and SWE-Bench Verified, respectively.

\textbf{Case study: sympy-12236.} The issue description states that a division of polynomials returned the wrong answer due to incorrectly simplifying the given expression. 

SWE-Agent looked for the \lstinline|apart| function and reproduced the issue using a test case. The agent's hypothesis was that the function was using integer division instead of field division during polynomial division and thus planned to modify the division logic. The patch added a check during the simplified integer-based calculation, comparing the simplified value to the original and retaining the original field type if they were not mathematically identical. This is a different approach from the gold patch, which targets a specific utility function to convert fractions to polynomial and adds a more robust case. The SWE-Agent patch displays less mathematical knowledge, but still achieves the desired fix using a more defensive programming approach.

Prometheus reproduced the error but did not start with the right hypothesis, instead focusing on symbolic values and modifying the logic around symbol handling in the n the \lstinline|apart| function in \lstinline|sympy/polys/partfrac.py|. Verifying this change led to recursion and recursion errors, which the agent attempted to fix using additional logic. It was not able to re-evaluate its hypothesis even though it entered multiple loops to reason about the error.

The reasoning of OpenHands was similar, reproducing the issue but identifying the problem to be with symbolic values in the \lstinline|apart| function. Like Prometheus, its attempted fixes led to other, increasingly unrelated error messages that it tried to patch with extra code. The agent generated a patch and ran test cases that produced the expected output, however, this patch was incorrect as it failed the unit test for the \lstinline|div| method. Thus, the root cause of the issue was missed. 

\begin{boxB}\textbf{Finding 3: Defensive programming can succeed even when agents lack deep domain knowledge, suggesting that robustness may compensate for limited problem-specific reasoning.}\end{boxB}

\subsection{RQ2: Length of Successful and Failing Trajectories}

Table \ref{tab:trajectory_stats} presents a statistical comparison of trajectory lengths for both benchmarks and all agents. We consider the overall step count, including all step types. Across all configurations, \textbf{failed trajectories are longer on average and have a wider distribution} than successful ones. However, this discrepancy varies between agents as well as between benchmarks. 
\begin{boxB} \textbf{Finding 4: Failed trajectories are longer and have a wider distribution than successful ones. Early stopping or trajectory monitoring could reduce wasted computation on doomed attempts.} \end{boxB}

\paragraph{Agent profiles}
In absolute terms, both OpenHands and SWE-agent exhibit around 20 more steps in failures in SWE-Bench Lite, while Prometheus takes 49 more. In verified, the difference is greater, with OpenHands having 44 more steps on average, SWE-agent having 22 more, and Prometheus having 74 more. However, as the agents have differing levels of verbosity, it is useful to compute the increase in trajectory length for failures as a percentage of the successful length. SWE-agent failed trajectories are 12.6\% longer in SWE-Bench Lite and 18.5\% longer in SWE-Bench Verified, while OpenHands failed trajectories are 31.0\% longer and 82.5\% longer in the respective benchmarks. Prometheus lies in the middle with its failed trajectories being 56.6\% longer in Lite and 50.7\% longer in Verified, both representing a significant divergence on par with OpenHands in Lite. As SWE-agent trajectories are always longer, usually around two to three times as many steps as OpenHands, there may be inverse correlation between the average successful trajectory length and the severity of failure.  This suggests that a more granular, incremental problem solving strategy involving many small steps requires more steps to achieve a successful outcome, but wrong steps are less catastrophic. 

SWE-Agent also has the most modest difference in standard deviation between success and failure, where the latter is only marginally higher. In contrast, the range of values for failure in OpenHands is much greater, with a standard deviation of 60.63 versus 46.60 in Lite, and for Verified the value for failures is double that of successes. Prometheus exhibits the most extreme variation in this metric; its standard deviation for failures is nearly three times that of its successes in Lite (100.79 versus 34.62) and remains drastically higher in Verified. We can thus infer that two kinds of failures are encountered, the first clustered around the median, which fail relatively quickly, and a second kind that continues for a long number of steps and behaves unpredictably.

\begin{figure}[tb]
    \centering
    \includegraphics[width=\linewidth]{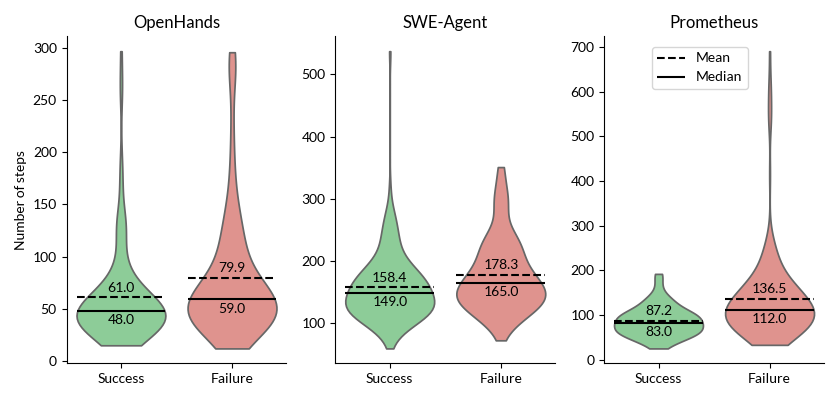}
    \caption{Trajectory Step Counts in SWE-Bench Lite}
    \label{fig:violin-lite}
\end{figure}

\begin{figure}[tb]
    \centering
    \includegraphics[width=\linewidth]{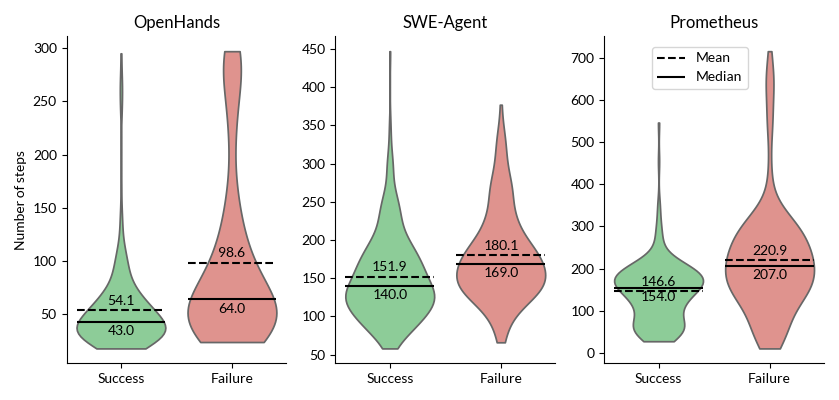}
    \caption{Trajectory Step Counts in SWE-Bench Verified}
    \label{fig:violin-verified}
\end{figure}

Figures \ref{fig:violin-lite} and \ref{fig:violin-verified} visualise the diverse and more drastic failures of OpenHands and Prometheus. OpenHands and Prometheus failures are longer tailed, with OpenHands having the same maximum length for failures and successes, and Prometheus having even longer failures. SWE-Agent fails faster, which is desirable for performance, resource efficiency, and learning ability. 
\begin{boxB} 
\textbf{Finding 5: A higher absolute number of steps in successful trajectories correlates with less divergent failures. Incremental reasoning styles (many small steps) may reduce catastrophic failures compared to short, aggressive approaches.}
\end{boxB}


\subsection{RQ3: Fault Localisation in Successful and Failing Trajectories}
\begin{table*}[tb]
\begin{ThreePartTable}
\renewcommand{\arraystretch}{0.95}
\centering
\small
\caption{Fault localisation (FL) Performance by File, Function, and Hunk Level}
\label{tab:faultloc}
\begin{tabular}{lllrrrrrr}
\toprule
& & & \multicolumn{3}{c}{\textbf{SWE-Bench Lite}} & \multicolumn{3}{c}{\textbf{SWE-Bench Verified}} \\
\cmidrule(lr){4-6} \cmidrule(lr){7-9}
\textbf{FL Level} & \textbf{Outcome} & \textbf{Match} & \textbf{OpenHands} & \textbf{SWE-agent} & \textbf{Prometheus} & \textbf{OpenHands} & \textbf{SWE-agent} & \textbf{Prometheus} \\
\midrule
\multirow{4}{*}{File} & \multirow{2}{*}{Success} & 0 & 6.5\% (8) & 4.1\% (7) & 2.7\% (2) & 6.4\% (17) & 7.8\% (26) & 4.2\% (8) \\
& & 1 & 93.5\% (116) & 95.9\% (163) & 97.3\% (73) & 93.6\% (248) & 92.2\% (307) & 95.8\% (183) \\
\cmidrule(l){2-9}
& \multirow{2}{*}{Failure} & 0 & 20.2\% (35) & 18.6\% (24) & 27.5\% (55) & 37.3\% (85) & 40.7\% (68) & 32.9\% (51) \\
& & 1 & 79.8\% (138) & 81.4\% (105) & 72.5\% (145) & 62.7\% (143) & 59.3\% (99) & 67.1\% (104) \\
\midrule
\multirow{4}{*}{Function} & \multirow{2}{*}{Success} & 0 & 73.4\% (91) & 66.5\% (113) & 78.7\% (59) & 72.1\% (191) & 72.1\% (240) & 68.1\% (130) \\
& & 1 & 26.6\% (33) & 33.5\% (57) & 21.3\% (16) & 27.9\% (74) & 27.9\% (93) & 31.9\% (61) \\
\cmidrule(l){2-9}
& \multirow{2}{*}{Failure} & 0 & 72.8\% (126) & 75.2\% (97) & 82.0\% (164) & 85.1\% (194) & 86.2\% (144) & 83.9\% (130) \\
& & 1 & 27.2\% (47) & 24.8\% (32) & 18.0\% (36) & 14.9\% (34) & 13.8\% (23) & 16.1\% (25) \\
\midrule
\multirow{6}{*}{Hunk} & \multirow{3}{*}{Success} & 0 & 86.3\% (107) & 82.4\% (140) & 88.0\% (66) & 81.1\% (215) & 85.3\% (284) & 79.1\% (151) \\
& & 1 & 4.0\% (5) & 5.9\% (10) & 2.7\% (2) & 6.4\% (17) & 5.1\% (17) & 8.4\% (16) \\
& & 0.5 & 9.7\% (12) & 11.8\% (20) & 9.3\% (7) & 12.5\% (33) & 9.6\% (32) & 12.6\% (24) \\
\cmidrule(l){2-9}
& \multirow{3}{*}{Failure} & 0 & 88.4\% (153) & 94.6\% (122) & 91.0\% (182) & 94.7\% (216) & 97.6\% (163) & 93.5\% (145) \\
& & 1 & 1.2\% (2) & 1.6\% (2) & 0.5\% (1) & 0.9\% (2) & 0.0\% (0) & 1.3\% (2) \\
& & 0.5 & 10.4\% (18) & 3.9\% (5) & 8.5\% (17) & 4.4\% (10) & 2.4\% (4) & 5.2\% (8) \\
\bottomrule
\end{tabular}
\begin{tablenotes}
    \small %
    \item \textbf{Legend:} Percentage (Count). Percentages are calculated per outcome. 0 = No Match; 1 = Match; 0.5 = Close Match within 5 lines (Hunk level only).
\end{tablenotes}
\end{ThreePartTable}
\end{table*}

Table \ref{tab:faultloc} displays the percentage and number of trajectories that achieved each of the three levels of fault localisation (FL), meaning that they modified the same files, functions, or hunks as the gold
patch for a given issue. We compare success and failure outcomes in every configuration.

Successful outcomes unsurprisingly have a high frequency of file level matches across the board, with over 90\% targeting the same files as present in the gold patch. This number drastically drops at the function level, with only about 27\% of successful patches modifying the same function as the gold patch. SWE-agent's successful attempts in SWE-Bench Lite are slightly better than the others at targeting the gold functions with a value of 33.5\%. This means that file level localisation is necessary but not sufficient; successful patches almost always need to find the right file, but not necessarily the exact function, implying that agents can still succeed with approximate edits. Correct fixes could potentially be made at various parts of the file, for instance by creating a new function or modifying a wrapper or child method of the gold function. Similarly, only a handful of successful patches targeted the exact hunks as the gold patches, but around 9-11\% of patches were within five lines. 

\begin{figure*}[tb]
    \centering
    \includegraphics[width=\linewidth]{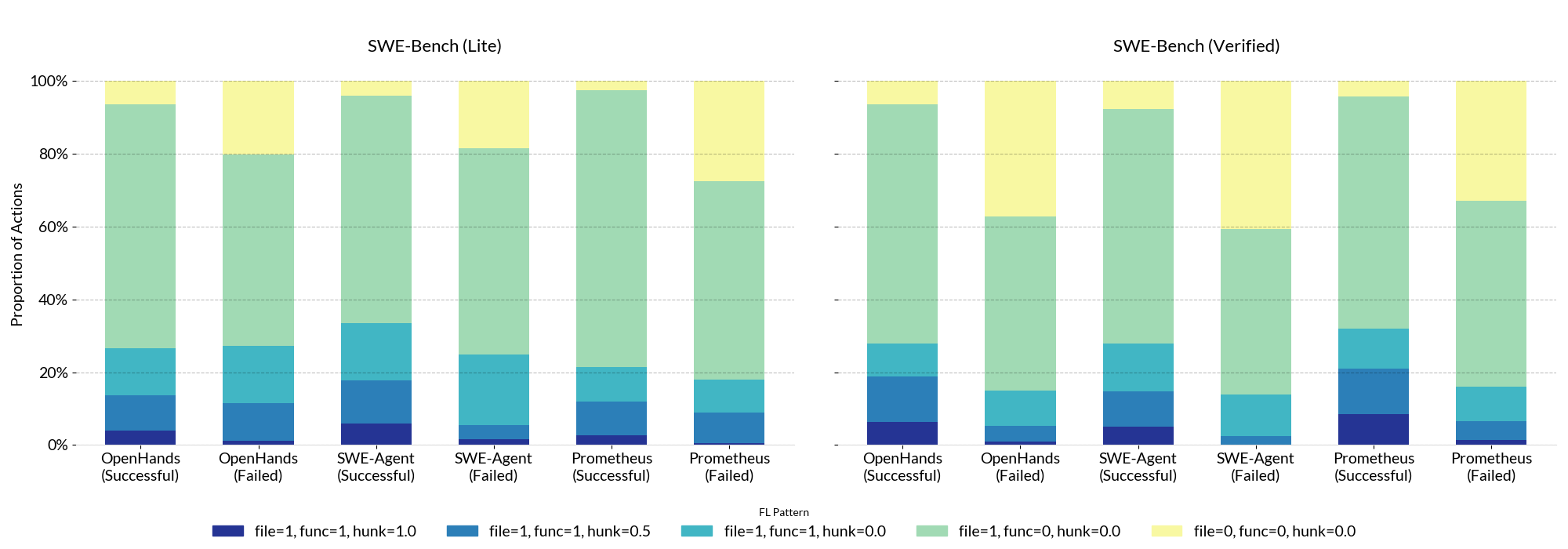}
    \caption{Proportion of Fault Localisation Combination Outcomes}
    \label{fig:faultloc_comb}
\end{figure*}

\begin{boxB}
\textbf{Finding 6: Successful patches find the correct patch file over 90\% of the time, but potential solutions within that file can be at multiple places.}
\end{boxB}

Failed outcomes also have a high frequency of file level matches, reaching as high as 81\% in the case of SWE-agent on SWE-Bench Lite. This highlights that \textbf{the majority of trajectories that fail are still able to locate the problematic file}, meaning that problems in fault localisation happen at a deeper level. Failed patches perform notably worse on SWE-Bench Verified, with 62.7\% and 59.3\% of patches targeting the gold file, compared to 79.8\% and 81.4\% in Lite. The file level again sees a drop in performance. The only outlier is OpenHands in SWE-Bench Lite, which achieves similar levels of function localisation for successful and failed patches, at around 27\%. In all other cases, failed patches are worse at targeting the gold functions than their successful counterparts, especially in Verified, where all three agents' failed attempts are only half as good. Notably, at the hunk level on the Lite benchmark, the performance of OpenHands' failed trajectories in producing close matches (10.4\%) is on par with that of all other successful ones (9.6\%–12.5\%). This pattern, observed at both the function and hunk level, indicates that partial progress towards a solution is not indicative of success for OpenHands on the issues in the Lite benchmark.
\begin{boxB}
\textbf{Finding 7: The majority of trajectories that fail are correctly able to locate problematic files. Failures often stem from fine-grained reasoning rather than coarse repository navigation.}
\end{boxB}

The frequencies of unique combinations of fault localisation outcomes at the file, function, and hunk levels are illustrated in Figure \ref{fig:faultloc_comb}. 
In all instances, the most frequent combination is file match, function miss, hunk miss, indicated as the largest component of each stacked bar, reinforcing the idea that trajectories are generally able to find the correct file, and spent most of their efforts exploring within the file. A complete failure pattern (file miss, function miss, hunk miss) seen at the top of every bar, is predominantly observed by failing trajectories and is more common in the Verified partition. At the file level, 35\% to 40\% of failed trajectories in Verified fail to locate the file, whereas only 20\% to 25\% of them fail in Lite. Recalling the findings from RQ2, where all agents had similar failing trajectory lengths in Verified and Lite, localising the wrong file does not correspond to a faster failure. This reveals that agents spend an equivalent amount of time exploring wrong and correct file paths in failed cases in SWE-Bench Verified. The small number of successful trajectories in this category indicates that solving bugs through code changes in another file is possible but unlikely.
\begin{boxB}
\textbf{Finding 8: In SWE-Bench Verified, failing solutions spend the same amount of time exploring wrong and correct file paths. Stronger signals for when to abandon unproductive paths are needed.}
\end{boxB}

\subsection{Discussion}

Prometheus' detailed context retrieval seems to help it solve complex issues, but sometimes causes it to get stuck in the search. Its failures were often due to reaching recursion limits, which we know to be high based on the trajectory lengths from RQ2. A direction for improvement can be thus to focus on abandoning paths and backtracking earlier. 

Although most SWE-Bench leaderboard submissions perform better on Verified than the other partitions, as the issues have been verified for solvability, RQ2 and RQ3 uncover surprising failure patterns. In RQ2, OpenHands and Prometheus trajectory length distributions dramatically vary in Verified compared to Lite; successes are more tightly clustered whereas failures are more widely spread out and unpredictable. SWE-agent is not affected by these changes. In RQ3, Verified contains a higher proportion of patches generated that failed to even find the correct file. These findings hint at the possibility that a higher clarity of tasks to solve may expose weaknesses in some agents' architecture that get hidden in the noisier Lite dataset. A more comprehensive inspection of results across both partitions, incorporating agent architectural features, is crucial to explore this possibility. 

Analysis of fault localisation reveals distinct areas for improving performance. For all agents, localising a fault at the hunk level, even within five lines of the gold patch, is a predictor of success. On SWE-Bench Verified, there remains significant room for improvement in file-level localisation, which is a bottleneck for a successful outcome. In contrast, on SWE-Bench Lite, where file localisation is already high (at least 72\% even for failed trajectories), there are only marginal gains from further improving this metric. In all cases, even successful trajectories rarely achieve a perfect line-level match, suggesting that aiming for a flawless match is an unproductive goal. Instead, enhancing the agent's capacity for close proximity hunk localisation represents a promising avenue for future development. Additionally, Finding 8 posits a potential avenue to achieve faster failures by detecting and abandoning unproductive trajectories, such as earlier detection of an incorrect file.

\section{Threats to Validity}\label{sec:threats}
The case studies in RQ1 cover only a small subset of trajectories, which may limit generalisability. We use uniform sampling to reduce bias, but the size of the sample may miss some behavioural diversity. To balance this, we combine in-depth qualitative analysis with large-scale quantitative results from RQ2 and RQ3. The extraction of thought, action, and result steps may be subjective. Extending to other agents requires consistent parsing across formats. To mitigate this, we focus only on repository exploration and code modification events, excluding implementation-specific ones, and make cross-agent comparisons mainly in relative terms.

\section{Conclusion}\label{sec:conclusion}
Our study advances the understanding of autonomous software engineering by moving beyond binary success metrics and introducing trajectory analysis as a key lens for evaluating code agents. Through a comparative study of OpenHands, SWE-agent, and Prometheus on the SWE-Bench benchmark, we reveal that agent trajectories encapsulate behavioural insights often obscured by outcome-only evaluation. Our findings show that agent success depends not merely on obtaining correct solutions, but on how agents gather context, recognise architecture patterns, abandon unproductive searches, and achieve approximate rather than exact fault localisation. We identify distinct problem-solving styles and failure modes, highlighting that longer, more variable trajectories often signal failure and that perfect fault localisation is neither necessary nor sufficient for success. These results highlight the need to move beyond leaderboard-based evaluation and develop frameworks for building more interpretable and robust code agents.

\bibliographystyle{ACM-Reference-Format}
\bibliography{Bibliography}

\appendix

\end{document}